Riccioli Measures the Stars: Observations of the telescopic disks of stars as evidence against Copernicus and Galileo in the middle of the 17th century

Christopher M. Graney
Jefferson Community & Technical College
1000 Community College Drive
Louisville, Kentucky 40272 (USA)
christopher.graney@kctcs.edu
www.jefferson.kctcs.edu/faculty/graney

G. B. Riccioli's 1651 *Almagestum Novum* contains a table of diameters of stars measured by Riccioli and his associates with a telescope. These telescopically measured star diameters are spurious, caused by the diffraction of light waves through the circular aperture of the telescope, but astronomers of the time, including Riccioli and Galileo Galilei, were unaware of this phenomenon. They believed that they were seeing the physical bodies of stars. In the *Almagestum Novum* Riccioli uses these telescopically measured disks to determine the physical sizes of stars under both geocentric (or geo-heliocentric – Tychonic) and heliocentric (Copernican) hypotheses. The physical sizes obtained under the Copernican hypothesis are immense – dwarfing the Earth, the Sun, and the Earth's orbit; even exceeding the distances to the stars given by Tycho Brahe. Thus Riccioli felt that telescopic observations were an effective argument against the Copernican system.

**Introduction**

G. B. Riccioli, in his 1651 *Almagestum Novum*, reports on observations of the telescopic disks of stars made by him and his team. The *Almagestum Novum* features a table of stellar disk diameters (measured telescopically), information on how the data in the table was collected, and conclusions drawn from the data about the distances and physical sizes[*] of stars. Like Tycho Brahe, who measured stellar disks with the naked eye decades earlier, Riccioli did not know that the disks he measured were spurious. And like Tycho Brahe, Riccioli used his telescopic observations of stellar disks to attack the Copernican hypothesis based on the immense physical sizes of stars required if they lay at the distances required by the absence of observable annual parallax. In the *Almagestum Novum* Riccioli also attacks the scientific integrity of Galileo Galilei, who Riccioli accuses of being deceptive in attempting to use telescopic stellar disk observations to support Copernicus. The *Almagestum Novum* shows that 17$^{th}$ century astronomers did not consider stars viewed telescopically to be mere points or blazes of light, but rather to be measurable bodies that might help settle "The Great Cosmological Controversy"[†] – and most likely in favor of a Tychonic geo-heliocentric world system.

**The Telescopic Disks of Fixed Stars**

Old books with language that seems a little exotic to today's reader sometimes tell stories from astronomy's history very well. This piece from an 1836 encyclopedia discussing one of Tycho's objections to the Copernican system is worth reading:

> The stars, to the naked eye, present diameters varying from a quarter of a minute of space, or less, to as much as two minutes. The telescope was not then invented which shows that this is an optical delusion, and that they are points of immeasurably small diameter. It was certain to Tycho Brahe, that if the earth did move, the whole motion of the earth in its orbit did not alter the place of the stars by two minutes, and that consequently they must be so distant, that to have two minutes of apparent diameter, they must be spheres as great a radius at least as the distance from the sun to the earth. This latter distance Tycho Brahe supposed to be 1150 times the semi-diameter of the earth, and the sun about 180 times as great as the earth. Both suppositions are grossly incorrect; but they were common ground, being nearly those of Ptolemy and Copernicus. It followed then, for any thing a real *Copernican* could show to the contrary, that some of

---

[*] I will use the term *physical* (physical size or physical diameter or physical radius) to refer to the physical extent of a star's globe as calculated by Riccioli or others. This is in contrast to the size or diameter or radius observed by Riccioli or others, either telescopically or via the naked eye.

[†] I borrow this phrase from the book by Harald Siebert (2006) of the same title.

> the fixed stars must be 1520 millions of times as great as the earth, or nine millions of times as great as they supposed the sun to be. Now, one of the strong arguments against Ptolemy (and the one which has generally found its way into modern works) was the enormous motion which he supposed the stars to have. The Copernican of that day might have been compelled to choose between an incomprehensibly great magnitude, and a similar motion. Delambre, who comments with brief contempt upon the several arguments of Tycho Brahé, has here only to say, 'We should now answer that no star has an apparent diameter of a second.' Undoubtedly, but what would you have answered *then*, is the reply. The stars were spheres of visible magnitude, and are so still; nobody can deny it who looks at the heavens without a telescope; did Tycho reason wrong because he did not know a fact which could only be known by an instrument invented after his death? ["Brahé, Tycho" 1836: 326]

This story nicely summarizes Brahe's objections to what Ann Blair (1990: 364) calls the "absurdity of the distance and the sizes of the fixed stars that the Copernican system required". It also summarizes that which is commonly accepted to be the answer to those objections – namely, that the telescope showed that, in the words of Stillman Drake (1957) –

> Fixed stars are so distant that their light reaches the earth as from dimensionless points. Hence their images are not enlarged by even the best telescopes, which serve only to gather more of their light and in that way increase their visibility. [47, note 16]

Because of this, Christine Schofield (1989) tells us that –

> The absolute size of stars was no longer a problem, because the use of the telescope had made necessary a re-estimation of the true diameter of the fixed stars, which were now known to be far smaller than they appear to the naked eye. [41]

All this is owed to Galileo's assessment in his 1610 *Starry Messenger* that –

> [T]he fixed stars are never seen to be bounded by a circular periphery, but have rather the aspect of blazes whose rays vibrate about them and scintillate a great deal. Viewed with a telescope they appear of a shape similar to that which they present to the naked eye, but sufficiently enlarged so that a star of the fifth or sixth magnitude seems to equal the Dog Star, largest of all the fixed stars. [Drake 1957: 47]

This notion that "the fixed stars appear as dimensionless points" has been repeated by the best of scholars from Kepler (Kepler and Wallis 1995: 46), who cites Galileo, to Van Helden (1985: 89), who cites Kepler, to Grant (1996: 448), who cites Van Helden.[*] Thus an article dedicated to

---

[*] The quote "the fixed stars appear as dimensionless points" are Van Helden's words. Kepler's quote is "Skilled observers deny that any magnitude as it were of a round body can be uncovered by looking through a telescope: or

scholastic reaction to Copernicanism in the 17th century (Grant 1984) or a book discussing Jesuit science during that time (Feingold 2003) will make no mention of a powerful anti-Copernican argument made by Riccioli, a well-known Jesuit, based on the fixed stars *not* appearing as dimensionless points.

Fixed stars do *not* appear as dimensionless points. Galileo did *not* view stars as points. As early as his letters on sunspots (1612/1613) Galileo was referring to stars as spheres:

> Stars, whether fixed or wandering, are seen always to keep the same shape, which is spherical. [Drake 1957: 100]

> [A certain gentleman] thinks it probable that even the other stars are of various shapes and that they appear round only because of their light and their distance, as happens with a candle flame – and, he might well have added, with horned Venus. Such an assertion could not be proven false if it were not that the telescope shows us the shapes of all the stars, fixed as well as planets, to be quite round. [Drake 1957: 137]

In fact, Galileo reports that stars are round in other works, including his 1617 observing notes on Mizar (Ondra 2004: 73-75), his 1624 letter to Ingoli (Galileo and Finocchiaro 1989: 167, 180), and his 1632 *Dialogue Concerning the two Chief World Systems* (359-360). His initial report on the stars, when he is a very inexperienced observer who has used a telescope for less than a year, says stars are "blazes" – after that he consistently reports them to be round.

Galileo reports stars to appear round because indeed stars seen through telescopes of small aperture and modest magnification (or even through larger telescopes at higher magnification) appear round, with brighter stars appearing larger than fainter stars. The round appearance, which may be accompanied by a series of progressively fainter concentric rings surrounding the apparent disk of the star, is owed to the star's image being in fact a classical "Airy pattern" created by the diffraction of light waves through the telescope's circular aperture, and to the limited sensitivity of the human eye (figure 1). George Biddell Airy fully described this phenomenon, which had been seen by astronomers from Galileo to Hevelius to Halley to Herschel,[*] in 1835:

> The rapid decrease of light in the successive rings will sufficiently explain the visibility of two or three rings with a very bright star and the non-visibility of rings with a faint star. The difference of the diameters of the central spots (or spurious disks) of different

---

rather, if a more perfect instrument is used, the fixed stars can be represented as mere points, from which shining rays, like hairs, go forth and are spread out."

[*] A full discussion of telescopic star disk observations over two centuries can be found in Graney and Grayson (2010).

stars (which has presented a difficulty to writers on Optics) is also fully explained. Thus the radius of the spurious disk of a faint star, where light of less than half the intensity of the central light makes no impression on the eye, is [smaller], whereas the radius of the spurious disk of a bright star, where light of 1/10 the intensity of the central light is sensible, is [larger]. [Airy 1835: 288]

While at the time of Airy astronomers understood these disks to be spurious, at the dawn of telescopic astronomy they did not. Galileo argues in the *Dialogue* (359-360) that stars of the first magnitude have diameters of 5 seconds of arc at most, and claims that this telescopically observed size, combined with the assumption that stars are suns, fully answers any objections to Copernicus arising from the absurdity of the distance and the physical sizes of the fixed stars:

…by assuming that a star of the sixth magnitude may be no larger than the sun, one may deduce by means of correct demonstrations that the distance of the fixed stars from us is sufficiently great to make quite imperceptible in them the annual movement of the earth…. [359]

He then proceeds to argue that a sixth-magnitude star appears $1/2160^{th}$ the diameter of, and so therefore is 2160 times more distant as, the sun. However, the argument does not work – at the star distances Galileo determines, the annual movement of the earth will be perceptible to an observer who follows Galileo's methods (Graney 2008). Only if the stars are true points will they not be a problem for Copernicus. This is argued by Simon Marius in his 1614 *Mundus Jovialis*. Marius, who also observed the telescopic stellar disks, states that they support the geocentric (or geo-heliocentric) hypothesis of Tycho Brahe (Graney 2010).

**Riccioli's *Almagestum Novum*, Chapter XI of Book 7, Section 6**

Galileo and Marius observed the telescopic disks of stars, so it should not be surprising that other $17^{th}$-century astronomers also observed them. Riccioli did, and includes information on telescopic stellar disks in his 1651 *Almagestum Novum*. The subject of this paper is his eleventh chapter (pages 715-717) of Section 6, Book 7, Volume I: This very small part of the *Almagestum Novum* illustrates the method by which Riccioli measured telescopic stellar disks, the type and amount of data he collected, and the conclusions he drew. It illustrates how telescopic observations of stellar disks were used to attack the Copernican hypothesis. It shows that the telescope did not answer Tycho's objection regarding the absurdity of the distance and the physical sizes of the fixed stars that the Copernican system required. More generally it illustrates that telescopic data could be used to argue for a geocentric world system – at least until astronomers discovered that the data was spurious. This discussion will simply follow Riccioli's Chapter XI.

In paragraph I, Riccioli begins by informing the reader that determining the apparent diameter of (fixed) stars depends upon observing planets and comparing planetary and stellar diameters; he refers the reader to Book 6 Chapter IX for discussion of the diameters of stars and planets and their measurement. He and Grimaldi record the shape of Jupiter's disk and Saturn's oval form, emphasizing that they can do this with accuracy by means of repeated and immediate comparisons between what they draw on paper and what they see through the telescope.[*] They confirm what their eyes show them by observing both together and separately, and by getting the opinion of an unbiased third party. Working during the turn of the year 1649/1650 they record the appearance of Jupiter and Saturn as shown in a figure in Chapter X of the *Almagestum Novum* (see figure 2). Saturn's diameter was 34''30''' and Jupiter's was 44''. They divide the diameter of Jupiter (BZ in figure 2) into 200 parts; the scale in the figure works out to 100 parts per Roman inch. Saturn's diameter FG on this scale is 160 parts.[†] Using planet drawings as references, they go on to observe stars, turning to look at the the drawings frequently to figure out the telescopic size of each star, measured in what is essentially hundredths of a Jovian radius. Then they get others to do this as well, especially P. Paulo Casato and P. Mattheo Taverna, who have "sharp eyes and minds". They are pleased to find close agreement in their various estimations. Riccioli says that anyone with a good telescope and a few unbiased observers can use this method to reproduce the results of his observing team. The telescopic star sizes the team obtained were recorded in units of hundredths of a Jovian radius, and calculated in terms of seconds and thirds of arc, and are seen in Table 1.

In paragraph II, Riccioli proceeds to comment on the telescopic star sizes observed by others – in particular by Hortensius and Gassendi, who measured the diameter of Sirius to be 10'' and the diameters of other first magnitude stars to be 8''. Riccioli blames the discrepancy on two things. The first is a matter of equipment and method – Hortensius had a poor telescope that did not magnify Jupiter much, and the process is sensitive to small errors of measurement and changes in the size of the image of Jupiter (owing to its location relative to Earth at any particular time). The second is a matter of bias – Hortensius, as a Copernican, is biased towards the stars being small. In the Copernican world system, the stars must be sufficiently distant that there is no perceptible annual parallax caused by the Earth's motion. Larger star sizes seen through the telescope translate, at those distances, into physically giant stars. Riccioli notes that the Copernican hypothesis with its moving Earth is far less supportable if the physical sizes of the stars dwarf the orbit of the Earth.[*] Riccioli also critiques Landsbergius who cites naked eye measurements of star sizes, including those of Tycho Brahe, that put the observed diameter of first magnitude stars at a minute (60'') or greater, but who then adds that through the telescope star diameters appear much smaller. Riccioli states that the telescope, in "exposing the disks of the stars and scraping off the adventitious ringlets of the rays" is more trustworthy than the naked eye or arbitrary estimation. Riccioli's use of the word "cincinnos" – ringlets or curls – is interesting. Riccioli is probably trying to describe the telescopic appearance of a bright star like Sirius, complete with diffraction rings (figure 1). If Riccioli is then adjusting his telescope to remove those rings – by restricting its aperture to the point where the rings are no longer detectable – then he is using a procedure that will tend to produce relatively uniform appearance in stars from one observer to another. Different observers using different aperture telescopes should see Sirius differently, with those observers using larger telescopes seeing a smaller spurous disk and more diffraction rings; but if all observers are adjusting their telescopes so that no rings are visible, then such differences are largely removed.

---

[*] In reading this, a sort of "blink comparison", in which one eye observes the paper while the other observes the image in the telescope, or a similar method, comes to mind.

[†] Note that {34.5/44}200 = 157. Riccioli also refers to Saturn's diameter as 35''. Note that {35/44}200 = 159. Riccioli's numbers are approximate, and, as the reader will see in Tables 1-4, full of rounding, typographical, and other editing errors.

In paragraph III, Riccioli turns his attention to Galileo. Riccioli remarks on how Galileo's 1632 *Dialogue* purports to answer objections to the Copernican system on account of the immense physical sizes of fixed stars required by the lack of observable parallax and the stellar diameters measured by the naked eye; Riccioli notes that even though Galileo that first magnitude stars do not exceed 5'' in diameter as seen through a telescope, Galileo still fails to solve the problem that he is claiming to answer (which Riccioli illustrates in Tables 2-4; note that he lists Alcor's telescopic diameter as being smaller than 5''). Riccioli also points out the inadequacy of Galileo's method of using suspended cords to verify star diameters measured telescopically. Riccioli is harsh in discussing Galileo: He seems to mock Galileo by echoing language from Galileo's discussion of this issue in the *Dialogue*; he uses terms like "fallax", "falsitate", and "falsa" both in the text of the paragraph and in the accompanying marginal note – these words convey that Galileo was not merely mistaken, but deceitful. Then Riccioli mentions Kepler, and how before the telescope Kepler said stars had large disks while after the telescope Kepler said the stars were but points – indeed, Riccioli uses the same quote from Kepler mentioned earlier in this paper. In closing the paragraph, Riccioli endorses following the evidence gathered by means of the telescope. He says he is providing Tables 1-4, which show the physical sizes of stars calculated based on telescopic observations and different estimates of distances to the stars (geocentric and Copernican), in order that they will be available for any discussion – and, Riccioli notes, especially so that they will be avaible for any discussion regarding the hypothesis of Copernicus. For in these tables, the old objection of Tycho is back in full force.  Under calculations based on a geocentric hypothesis, stars range in physical size from as small as a seventh of Earth's diameter to as large as 18 times Earth's diameter.  This puts stars in the general size category as the Earth, or the Sun, depending on the star and the details of the hypothesis used.  But calculations based on the Copernican hypothesis, especially those done based on lack of observed parallax, result in stars ranging in physical size from thousands to tens of thousands of Earth diameters.  Thus the stars dwarf the Earth, and even the Sun, being comparable in physical size to, or substantially larger than, the Earth's orbit; or even comparable to the distance to the stars themselves as calculated under a geocentric hypothesis. (Riccioli lists Tycho Brahe's distance to stars as 14,000 Earth radii; the physical sizes of a single star in the larger extremes of the Copernican hypothesis in Table 4 exceed this value.)  One can see why Riccioli wants these tables available for discussion.

---

[*] The difference between Riccioli's 18'' diameter for Sirius and the 10'' diameter he ascribes to Hortensius is not sufficient to make a difference in the argument, as Riccioli will implicitly note shortly in his criticism of Galileo. Riccioli probably overestimated the telescopic size of Jupiter (see figure 2c), so the difference between Riccioli's measurements and those of Hortensius is not that great.

     Riccioli finishes the chapter with a discussion of what must be done to get a telescope to rid a star image of its "adventitious rays", so that its disk can be seen and its true size revealed.  He says that the object lens of a telescope must be masked with foil or some other other thin material that has a hole in it of radius[*] approximately one quarter of a Roman inch.  Riccioli says that, with such an apparatus, a perfect, round disk will be seen which can be compared with the disks of Jupiter or Saturn.  Riccioli notes that the hole radius he gives is approximate, being larger for faint stars.  Thus it is indeed likely that Riccioli is masking his telescope so as to eliminate diffraction rings; such rings are fainter in fainter stars and will disappear with less reduction of a telescope's aperture (see figure 1).[†]

---

[*] Riccioli actually says "diameter" rather than "radius".  However, Riccioli's tables contain numerous errors where diameters are stated but radii given, and the telescopic star sizes he gives are more consistent with diffraction pattern calculations for an aperture of one quarter inch radius.  It seems probable that the same diameter/radius error exists here.

[†] The reader may question whether Riccioli is merely choosing an arbitrary aperture size:  Why stop the restriction of aperture once the rings are scraped off?  If restricting the aperture to ¼ inch radius renders a truer star image than less restriction, would not still further restriction render a still truer image?
     The reader should keep in mind the appearance of a star image when a refracting telescope is being focused.  When a star is optimally focused, its image is as small and intensely bright as possible.  Further adjustment in either direction (either lengthening or shortening the distance from object lens to eye lens) will result in the image swelling into a larger and less intense disk of light.  Now consider aperture restriction.  Owing to the phenomenon of diffraction, restriction of a telescope's aperture enlarges the spurious disk, but reduces its intensity (see Airy 1835).  In general, aperture restriction beyond that needed to eliminate diffraction rings results in the star image becoming a larger and less intense disk of light.  Thus an astronomer such as Riccioli probably viewed aperture restriction as a sort of secondary "focusing" for stars – one restricts the aperture until no rings are seen, but no further so as not to enlarge and apparently defocus the disk.  Because there is a fixed ratio between the intensities of the rings and the peak of the central disk, various observers should reach fairly consistent results when observing a bright star such as Sirius, differing mostly in when they judge the last ring to have been fully "scraped off".
     However, the interplay between aperture, intensity, disk and ring sizes, and the limits of sensitivity of the eye (which has two different types of light detecting cells of greatly differing sensitivies) is complex.  Faint stars such as Alcor will show no rings at all in a small telescope – how did Riccioli treat them?  A full treatment of the subject of what affects disk size, in light of Airy's paper, is beyond the scope of this paper but is largely available in

**Conclusion**

In conclusion, we see that Riccioli and his associates telescopically observed and measured the disks of stars. Understandably misinterpreting these as representing the physical extent of stars, Riccioli draws conclusions concerning the distances and physical sizes of stars; conclusions which show that, under the Copernican hypothesis, the stars would have to be comparable in physical size to the Earth's orbit, or larger. Riccioli uses these to attack the Copernican hypothesis via the same argument as Tycho Brahe had used, and has harsh things to say about Galileo, who he says has been deceptive on this issue.

This line of attack would have to eventually collapse. At the time Riccioli's team was making their observations, Jeremiah Horrocks had already observed lunar occultations of stars, noting that the stars disappeared instantaneously. Edmund Halley (1720) would use such occultation data to question Cassini's interpretation of the telescopic disk of Sirius as representing its physical extent. However, until the spurious nature of telescopic disks was uncovered, the telescope would appear to support a Tychonic world system, not a Copernican one (Graney 2010: 19-22).

A review of Edward Grant's paper on the scholastic reaction to Copernicanism in the 17th century once characterized a thorough study of Riccioli's defense of Copernicanism as being "too thankless a task" to undertake, for work such as the *Almagestum Novum* is "sometimes tedious or even apparently stupid" (Eastwood 1985). Hopefully this paper will lead others to further investigate the question of telescopic observations of stars and the role they played in "The Great Cosmological Controversy", and to bury the myth that work such as Riccioli's is tedious or stupid. No doubt, there is more on telescopic stellar disks waiting to be unearthed.[*]

---

Graney (2009) and Graney and Grayson (2010). The purpose of this note is only to illustrate that Riccioli's statement that there is an approximate optimum aperture for viewing the disk of a star, scraped of its adventitious rays, is generally consistent with what is understood concerning star images seen through a small aperture telescope.

[*] A final footnote to this topic: Riccioli's observations of stars have not completely escaped notice. Juan Casanovas (1985) mentions that Riccioli measured the telescopic diameter of Sirius and made stellar distance and physical size calculations based on parallax (68). Casanovas even discusses how Galileo believed "he could see the real disk of a star [72]." However, Casanovas cites Kepler (again the same quote as previously mentioned), and concludes that "it was realized that the angular diameters of stars were not perceptible with telescopes, which were seen to be still very imperfect [73]." Early telescopes could be far from imperfect (Greco, Molesini, Quercioli 1992); for example, Galileo and Hevelius each produced very precise stellar observations with them (Graney 2007; Graney 2009). In *The Assayer* Galileo defends the validity of telescopic views of stars, attributing any possible minor differences between a telescope's view of the moon and its view of the stars to merely changes in the required focus (Drake 1957).

**Acknowledgements**

First, I must acknowledge my wife, Christina Graney.  Without her brains and her language skills ("That's a future imperfect conjunctive verb participle phrase so it has to go with this adverbial genitive clause in ablative case!") it would have been a *very* long time before I figured out Riccioli's writing – if I ever did.  I wish to acknowledge the HASTRO-L History of Astronomy discussion listserver and its members, including John McMahon, Giancarlo Truffa, Michael Crowe, Owen Gingerich, Jim Lattis, Michael Shank, and Steve McClusky who all offered some commentary on this project; knowing that other people found it interesting motivated me to find out what Riccioli was saying.  Language translation software was provided by Otter Creek-South Harrison Observatory, whose budget is provided by Jefferson Community & Technical College of Louisville (Kentucky, USA) and Harrison County (Indiana, USA) Parks & Recreation.

**Full text of *Almagestum Novum* Book 7, Section 6, Chapter XI**

CAPUT XI

*Nostrae et Aliorum Observationes circa Diametros Apparentes Fixarum, ex Planetarum Diametris deductas, cum Vera Fixarum Magnitudine, tam secundum nostras, quam secundum aliorum distantiam Fixarum a terra vel assertam vel asserendam.*

I. Libro 6. cap. 9. ob dependentiam, quam habet observatio diametri Fixarum ab observatione diametri Planetum, et affinitatem evidentiae, distulimus in hunc locum Fixarum observationes circa diametros apparentes factas, et hinc post Jovialis ac Saturnalis diametri certam definitionem postremo confirmatas. Potissimus et omnium evidentissimus modus hic fuit. Descripsimus in candida papyro varias imagines tum circulares pro disco Jovis, tum oviformes pro Saturno com comitibus ipsius; et inspecto diu multumq. utroque Planeta per grande Telescopium, elegimus ex multis illam figuram, que oculo a Telescopio ad papyrum statim ac identidem traducto, visa est omnium congruentissima quantitati disci Planetae tunc apparentis. Ne vero nostris tantum oculis fidem praeberemus, P. Grimaldus et ego tum simul tum seorsim adhibuimus plurium aliorum oculos contestes, qui nihil praemoniti de figura a nobis electa, eligeret ipsimet eam, quae congruentissima videbatur. Itaque electa fuit sub finem Anni 1649 et initium 1650 illa Saturni, et Jovis, quam exhibuimus cap. 10 num. 8 erat autem tunc, ex dictis ibidem, Saturni diameter apparens 34''30''' et Jovis 44''. Divisimus autem diametrum BZ, Jovialis disci in particulas 200. qualium uncia pedis Romani est 100. et talium partium fuit interior seu solius Saturni diameter FG 160. His praeparatis: rursus descripsimus in charta complures circellos majoris ac minoris diametri, et inspectis Fixis stellis, de quibus infra, per iem Telescopium, ac retorquendo frequenter oculum ad illos circellos, elegimus illum, qui quantitati stellae Fixae visus est aequalissimus omnium: Deindeo seorsim adhibitas alijs et in primis P. Paulo Casato, et P. Mattheo Taverna, illo Philosophiae; hoc sacrarum concionum eximijs professoribus, et ob acumen oculi ac mentis, atque acrimoniam iudicij aptissimis ad haec discernenda; qui eodem Telescopio usi, mira sane conspiratione, singuli elegerunt illas ipsas figuras, quas P. Grimaldus et ego elegeramus. Itaque sublata omni formidine et suspentione, quae nos antea tenuerat, tanquam de stellis tam nobili piscatu captis laetati; diametros illorum circellorum acutissimo circino comprehensas applicuimus diametro Jovis et Saturni, et ex proportione, quam habebant cum illis, elicuimus quantitatem apparentem diametri infrascriptatum Fixarum, eique securissime acquieuimus: nec dubitamus si omnia praedicta experimenta a peritis viris repetantur, nec illi alienis de Planetarum quantitate aut distantia Fixarum opinionibus imbuti sint, utanturque Tubospicillo et testibus idoneis, quin absque sensibili discrimine consensuti sint nobiscum.

II. Nam scimus quidem Hortensium in dissertatione cum Gassendo, valde diminuiste Fixarum diametros, adeo ut Sirio 10''. tantummodo, reliquis primi honoris stellis 8''. concesterit; sed duabus de causis: primo quia comparado earum diametrum cum Jovis diametro, usus est

Telescopio nimis parvam Jovis imaginem exhibente, ut patet ex schematismo imaginis ab ipso tradito, de quonos in scholio 4. capitis 10. eius enim diameter erat tantummodo partium 60. qualium nostri Jovis diameter apogeo vicini erat 200. ut in ipsa porportione inquirenda facilius periculum suerit errandi in minori, quam in majori mensure:  Secundo, quia totus erat in defendenda hypothesi Copernicana, quoad distantiam Fixarum in ea necessarium, ne parallaxis Fixarum ex motu annuo telluris orta, sensibilis evadat; Quia vero ex tanta distantia, sequitur magnitudo Fixarum nimia, et plerisque visa enormis, ut eam criminationem dilueret, diluendas putauit Fixarum diametros et magnitudines.  Id vero colligo, tum quia concludit: *Non sequitur ergo ex Coperinicao motu terrae, Fixarum magnitudinem, in immensum augeri, cum videamus eas esse multo minores annuo orbes terrae contra quosdam;* tum ex hoc ipso quasi astu, quo et ipse et Landsbergius non expresserunt Fixarum quantitatem comparatam cum globo Terrae, ne nimia videretur, sed cum sphaera orbis annui Terrae.  Quamuis *Landsbergius* in Uranometria lib. 3.clemento 20. nec Hortensium nec alios secutus, pro mero arbitrio maiusculam ponat diametrum Fixarum his verbis: *Semidiametrum apparentem Stellarum inerrantium primae magnitudinis, Albategnius definit 45''. Et Tycho Braheus uno scrupulo primo: Nos scrupulo primo dimidio. Nam per Tubum Opticum apparet adhue multo minor.*  Quasi vero non plus credendum fit Tubo denudanti discos stellarum, et abradenti cincinnos radiorum adventitios, quam nudo oculo, aut mero arbitrariae aestimationis iudicio.

III.  Simili suspicione, qua Hortensius, laborat apud nos *Galileus, et Keplerus*.  Nam Galilaeus dialogo 3. de systemate Cosmico respondens ijs, qui Copernicae hypothesi obiecerant immensitatem Fixarum ex ea consurgentem, recte quidem docet, ante usum specilli Belgici, nimis excessisse Astronomos in quantitate diametri apparentis minorum Planetarum, et Fixarum; sed ipse tamen nimis ab eadem deficit, dum negat stellas primae magnitudinis excedere 5''. secunda Scrupula; ac proinde non sequi, ut illae sint majores orbe annuo Telluris.  Potest enim facile a quovis, adhibere tubos Galilaei Telescopio grandiores ac perfectiores, evinci de falsitate.  Quam vero fallax fit modus observandi harum diametros per filum suspensum, satis docuimus lib. 6. cap. 9. num. 3. *Keplerus* quoque qui in libro de stella nova cap. 16. & 21. ante usum Telescopij, conceslerat Sirio 4'. minuta, et cuius de cingulo Orionis 2'. saltem; postea in Epitome Astronomiae pag. 498. quaestiunculae illi, Quantae appareant Fixae ex tellure visae? respondet: *Periti artifices negant ullam quantitatem veluti rotundi corporis, detegi per inspectionem Telescopij, quia potius quo perfectius instrumentum, hoc magis Fixas repraesentari ut puncta mera, ex quibus radij lucidi in speciem crinium exeant, disperganturque.*  Ibidem quoque existimat, Solem majorem esse quoad molem corporis, quam Fixas.  Virum, his praetermissis, sequemur evidentiam observationum grandi et praevalido Telescopio a nobis peractarum, et subijciemus Tabulas sequentes, cum Magnitudine vera, ex varijs distantijs, eruta, ut sint paratae in omnem disputationis eventum, praesertim pro hypothesi Copernicana.

Quod autem periti artifices, de quibus modo Keplerus, Fixas viderint ceu puncta radiantia, id si non ex imperfectione oculorum, ex modo tamen Telescopij adhibendi evenisse putandum est: nam si lenti ab oculo remotiori superposuissent bracteam seu laminam in medio perforatam, foramine, cuius diameter aequet quartam circiter partem unciae pedis Romani; vidissent Fixas ita radijs adventitijs detonsas, ut et perfecte rotundae apparent, et non ut puncta, sed disculi adeo vi Telescopij ampliati, ut certam quantitatem exhiberent, comparanti eos cum disco Jovis aut Saturni, notae modo quo supra diximus quantitatis. Dixi *circiter*, quia foramen illud in stellis minoribus, maius; in majoribus minus aliquanto esse debet: ut experienti constare poterit.

**Figures**

Figure 1

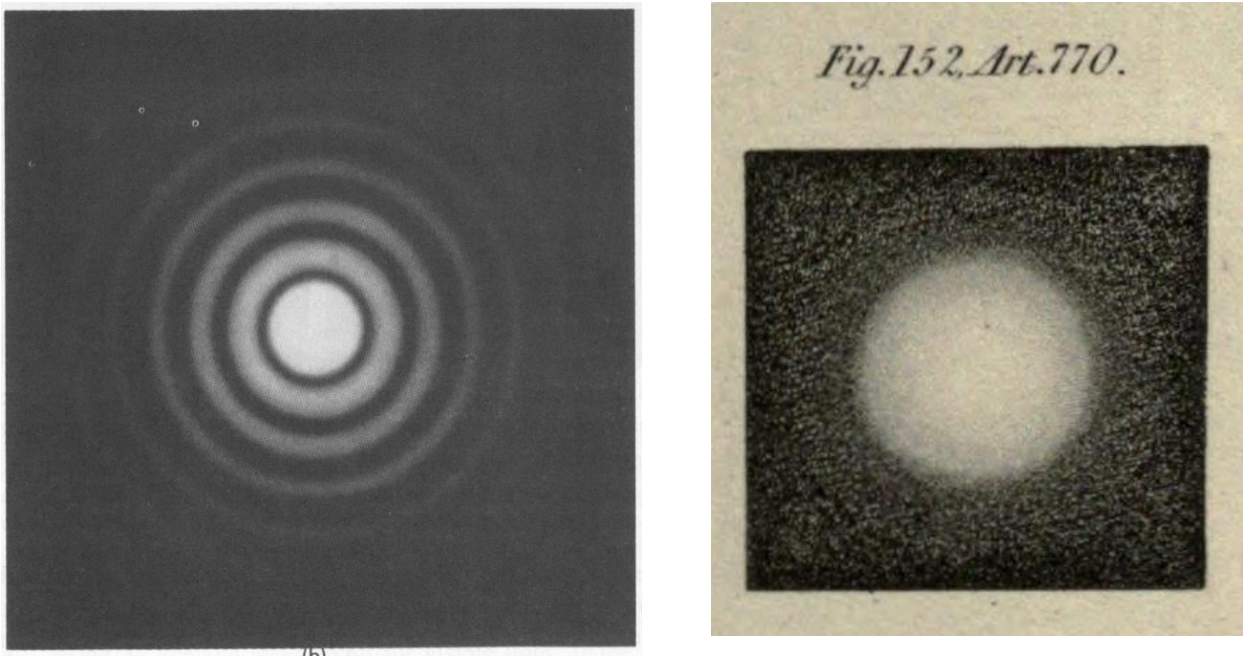

Figure 1: A bright star seen through a telescope with a small aperture and modest magnification shows a distinct disk and several faint diffraction rings (left) – the classic diffraction pattern formed by waves of light passing through the telescope's circular aperture. These rings are less visible, and even invisible, for fainter stars, or if the telescope's aperture is reduced. John Herschel (1828: 491-492) provides an excellent description of these disks which is useful for understanding the varying reports of 17$^{th}$-century astronomers:

> When we look at a bright star through a very good telescope with a low magnifying power, its appearance is that of a condensed, brilliant mass of light, of which it is impossible to discern the shape for the brightness; and which, let the goodness of the telescope be what it will, is seldom free from some small ragged appendages or rays. [But under greater magnification] the star is then seen … as a perfectly round, well-defined planetary disc, surrounded by two, three, or more alternately dark and bright rings, which … succeed each other nearly at equal intervals round the central disc.... [T]he apparent size of the disc is different for different stars, being uniformly larger the brighter the star.

Herschel discusses how reducing the aperture of the telescope to approximately one half of an inch makes the rings invisible, and the disk large and prominent. He provides a representation of the star's appearance through such an aperture (right).

Figure 2

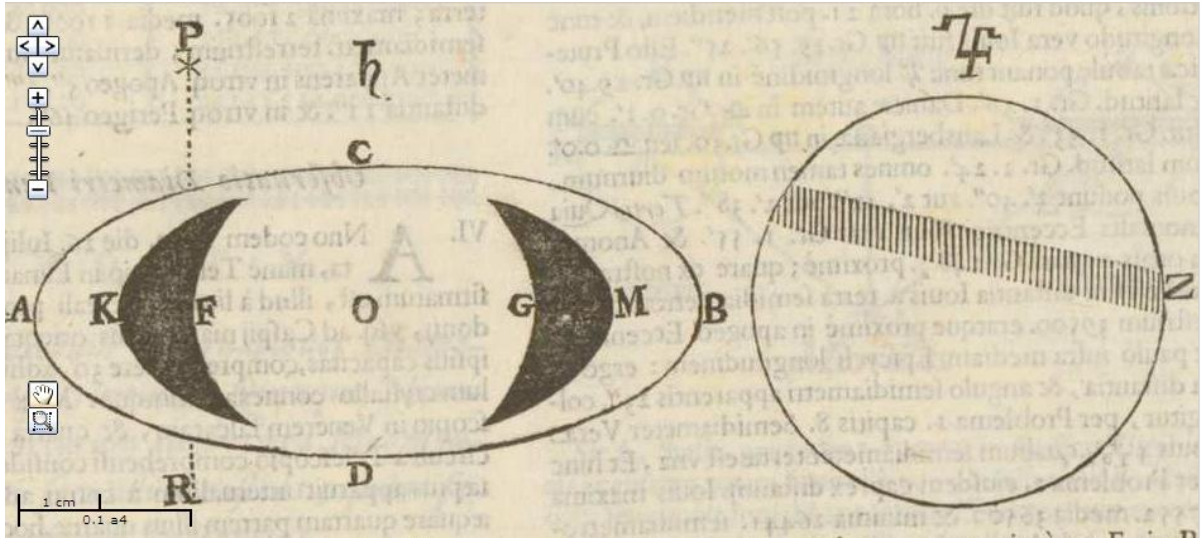

Figure 2A: Riccioli's Jupiter and Saturn figures.

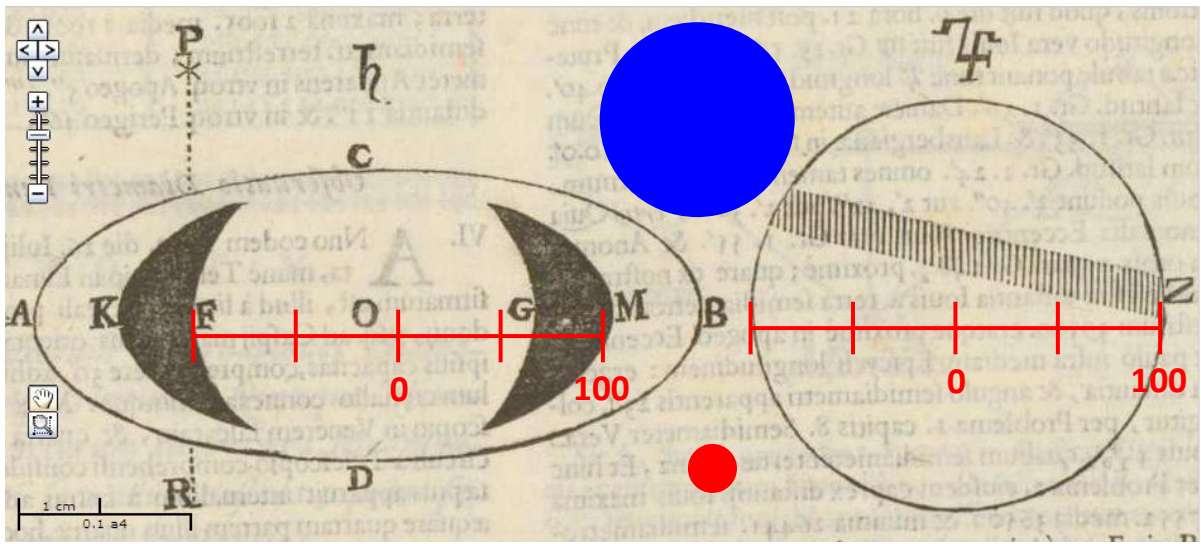

Figure 2B: Riccioli's Jupiter and Saturn figures showing scale of 1 Jovian Radius = 100 units. Also shown for purpose of comparison are representations of the sizes of Sirius (blue disk) and Alcor (red disk) as given by Riccioli in his tables.

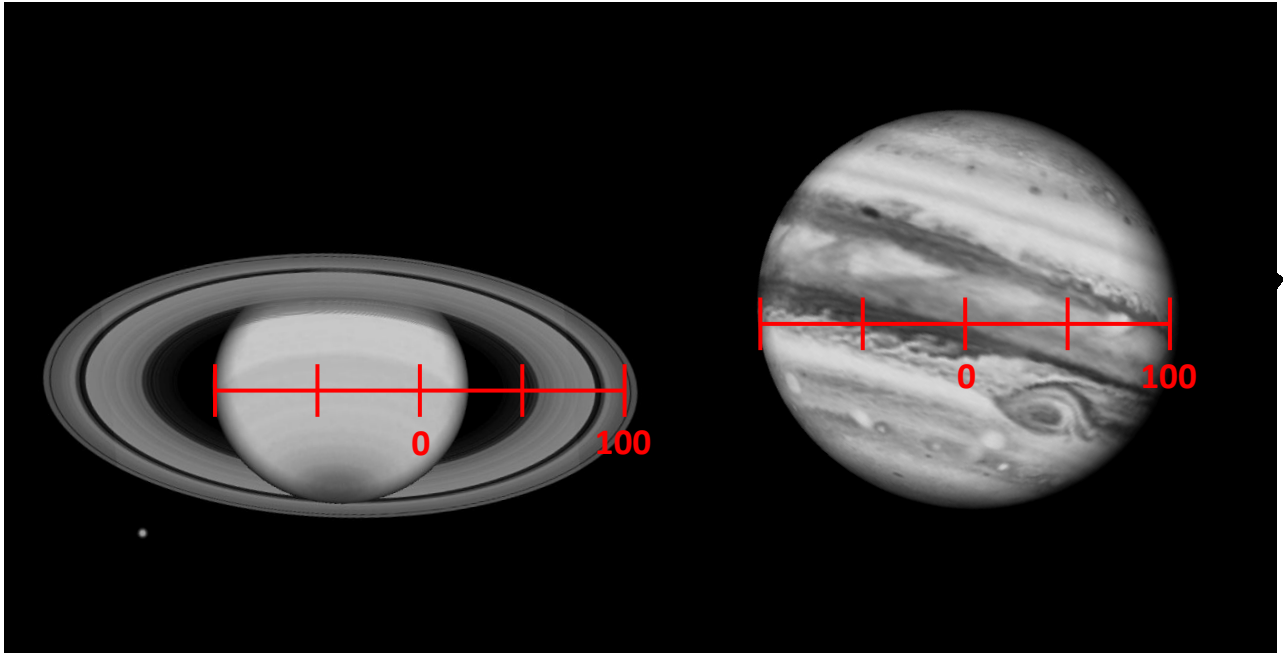

Figure 2C: Representations of Jupiter and Saturn for January 1, 1650 produced using the computer program *Stellarium*, showing their relative sizes and Saturn's general appearance. Compare to Riccioli's diagrams. According to *Stellarium*, Jupiter's apparent size at the time was 34''.

| TABLE 1: Diameters of the fixed stars, determined by way of comparison to the disk of Saturn and Jupiter when those disks measured 35" and 44" respectively. | | | | | | | |
|---|---|---|---|---|---|---|---|
| A | B | C | D | E | F | G | H |
| Fixed star | Diameter (200 units per Jovian diameter) | Apparent diameter (seconds and thirds of arc) | | Ancient magnitude ranking | Apparent diameter (seconds and thirds of arc), calculated from column B. | | Diameter (200 units per Jovian Diam.), calculated from columns C and D |
| Sirius | 82 | 18 | 0 | 1 | 18 | 2 | 81.82 |
| Lyrae lucida | 79 | 17 | 24 | 1 | 17 | 23 | 79.09 |
| Arcturus | 76 | 16 | 42 | 1 | 16 | 43 | 75.91 |
| Capella | 73 | 16 | 8 | 1 | 16 | 4 | 73.33 |
| Aldebaran | 70 | 15 | 24 | 1 | 15 | 24 | 70.00 |
| Spica | 68 | 15 | 5 | 1 | 14 | 58 | 68.56 |
| Regulus | 64 | 14 | 5 | 1 | 14 | 5 | 64.02 |
| Regel | 62 | 13 | 40 | 1 | 13 | 38 | 62.12 |
| Fomahant | 61 | 13 | 25 | 1 | 13 | 25 | 60.98 |
| Antares | 60 | 13 | 12 | 1 | 13 | 12 | 60.00 |
| Hydra | 58 | 12 | 45 | 1 | 12 | 46 | 57.95 |
| Cauda Serp. | 57 | 12 | 30 | 1 | 12 | 32 | 56.82 |
| Procyon | 56 | 12 | 20 | 2 | 12 | 19 | 56.06 |
| Aquila | 50 | 11 | 0 | 2 | 11 | 0 | 50.00 |
| Orion. cingul. | 40 | 8 | 50 | 2 | 8 | 48 | 40.15 |
| Coronae lucida | 38 | 8 | 21 | 2 | 8 | 22 | 37.95 |
| Polaris | 36 | 7 | 54 | 2 | 7 | 55 | 35.91 |
| Medusae Caput | 32 | 7 | 3 | 3 | 7 | 2 | 32.05 |
| Propus | 28 | 6 | 10 | 4 | 6 | 10 | 28.03 |
| Pleias lucidior | 24 | 5 | 16 | 5 | 5 | 17 | 23.94 |
| Alcor | 20 | 4 | 24 | 6 | 4 | 24 | 20.00 |

Table 1: Riccioli's telescopic star size data table (highlighted columns) from the *Almagestum Novum*. Calculated values are included for comparison, illustrating rounding errors in the data. It seems likely that the values given in column B have been rounded from original values used to calculate columns C and D.

| A | B | C | D | E | F | G |
|---|---|---|---|---|---|---|
| TABLE 2: True size of largest fixed star (Sirius), and the smallest (represented by Alcor): based on the diameters we determined for Sirius and Alcor of 18" and 4"24''', respectively, and on the distance to the stars in the hypothesis of a "resting Earth" | | | | | | |
| | | | TRUE SIZE OF SIRIUS | | TRUE SIZE OF ALCOR | |
| | | Distances of stars from Earth, measured in Earth radii | Diameter (in Earth diameters) | Volume (in Earth volumes) | Diameter (in Earth diameters) | Volume (in Earth volumes) |
| Tycho | | 14000 | 0.61 | 0.2 | 0.15 | 0.0003 |
| Ptolemaici | Max. | 40000 | 3.5 | 42 | 0.86 | 0.7 (?) |
| (Riccioli) | | 210000 | 17.5 | 5355 | 4.0 | 64.0 |
| *With volumes calculated from Riccioli diameter values:* | | | | | | |
| Tycho | | 14000 | 0.61 | 0.2 | 0.15 | 0.0034 |
| Ptolemaici | Max. | 40000 | 3.5 | 43 | 0.86 | 0.6 |
| (Riccioli) | | 210000 | 17.5 | 5359 | 4.0 | 64.0 |
| *With diameters and volumes calculated from the given distances of stars from Earth, and using the provided diameters of Sirius and Alcor:* | | | | | | |
| Tycho | | 14000 | 0.61 | 0.2 | 0.15 | 0.0033 |
| Ptolemaici | Max. | 40000 | 1.75 | 5 | 0.43 | 0.1 |
| (Riccioli) | | 210000 | 9.16 | 769 | 2.2 | 11.2 |
| *With diameters and volumes calculated as above but in units of Earth radii:* | | | | | | |
| | | | Diameter (in Earth radii) | | Diameter (in Earth radii) | |
| Tycho | | 14000 | 1.22 | 1.8 | 0.30 | 0.0266 |
| Ptolemaici | Max. | 40000 | 3.49 | 43 | 0.85 | 0.6 |
| (Riccioli) | | 210000 | 18.33 | 6155 | 4.5 | 89.9 |

Table 2: Riccioli's table (highlighted) from the *Almagestum Novum* of calculated physical diameters and volumes of stars, compared to the Earth, based on distances to the stars provided by various geocentric astronomers. Calculations based on Riccioli's numbers are included for purpose of comparison. These illustrate rounding errors, factor-of-two errors in the Ptolemaici and Riccioli figures (apparently calculated in Earth radii rather than the stated Earth diameters), and a factor-of-ten error in the Tycho values (column G).

| A | B | C | D | E | F | G |
|---|---|---|---|---|---|---|
| TABLE 3: True size of largest fixed star (Sirius), and the smallest (represented by Alcor): based on the diameters we determined for Sirius and Alcor of 18" and 4"24''', respectively, and on the distance to the stars claimed by supporters of the Copernican hypothesis. | | | | | | |
| | | | TRUE SIZE OF SIRIUS | | TRUE SIZE OF ALCOR | |
| | | Distances of stars from Earth, measured in Earth radii | Diameter (in Earth diameters) | Volume (in Earth volumes) | Diameter (in Earth diameters) | Volume (in Earth volumes) |
| Hortensius | | 10,312,227 | 899 | 726,572,600 | 442 | 86,355,888 |
| Galilaeus | | 13,046,400 | 1138 | 1,473,760,072 | 558 | 173,741,112 |
| Landsbergius | | 41,958,000 | 3658 | 48,947,466,312 | 1796 | 5,793,206,336 |
| Keplerus | | 60,000,000 | 5232 | 143,219,847,228 | 2568 | 16,933,994,432 |
| *With volumes calculated from Riccioli diameter values:* | | | | | | |
| Hortensius | | 10,312,227 | 899 | 726,572,699 | 442 | 86,350,888 |
| Galilaeus | | 13,046,400 | 1138 | 1,473,760,072 | 558 | 173,741,112 |
| Landsbergius | | 41,958,000 | 3658 | 48,947,566,312 | 1796 | 5,793,206,336 |
| Keplerus | | 60,000,000 | 5232 | 143,219,847,168 | 2568 | 16,934,994,432 |
| *With diameters and volumes calculated from the given distances of stars from Earth, and using the provided diameters of Sirius and Alcor:* | | | | | | |
| Hortensius | | 10,312,227 | 450 | 91,098,143 | 110 | 1,330,608 |
| Galilaeus | | 13,046,400 | 569 | 184,469,342 | 139 | 2,694,416 |
| Landsbergius | | 41,958,000 | 1831 | 6,136,157,024 | 448 | 89,626,612 |
| Keplerus | | 60,000,000 | 2618 | 17,943,447,153 | 640 | 262,087,552 |
| *With diameters and volumes calculated as above but in units of Earth radii:* | | | | | | |
| | | | Diameter (in Earth radii) | | Diameter (in Earth radii) | |
| Hortensius | | 10,312,227 | 900 | 728,785,141 | 220 | 10,644,862 |
| Galilaeus | | 13,046,400 | 1139 | 1,475,754,734 | 278 | 21,555,331 |
| Landsbergius | | 41,958,000 | 3662 | 49,089,256,192 | 895 | 717,012,894 |
| Keplerus | | 60,000,000 | 5236 | 143,547,577,224 | 1280 | 2,096,700,415 |
| *With diameters and volumes calculated as above but in units of Earth radii with factor-of-two error on Alcor:* | | | | | | |
| | | | Diameter (in Earth radii) | | Diameter (in Earth radii) x 2 | |
| Hortensius | | 10,312,227 | 900 | 728,785,141 | 440 | 85,158,894 |
| Galilaeus | | 13,046,400 | 1139 | 1,475,754,734 | 557 | 172,442,649 |
| Landsbergius | | 41,958,000 | 3662 | 49,089,256,192 | 1790 | 5,736,103,154 |
| Keplerus | | 60,000,000 | 5236 | 143,547,577,224 | 2560 | 16,773,603,317 |

Table 3: Riccioli's table (highlighted) from the *Almagestum Novum* of calculated physical diameters and volumes of stars, compared to the Earth, based on distances to the stars provided by various Copernican astronomers. Calculations based on Riccioli's numbers are included for purpose of comparison, illustrating errors much like those found in Table 2, and also additional factor-of-two errors in the values for Alcor.

TABLE 4: True size of largest fixed star (Sirius), and the smallest (represented by Alcor): based on the diameters we determined for Sirius and Alcor of 18" and 4"24"', respectively, and on the distance to the stars required by the Copernican hypothesis if annual parallax of the fixed stars is not to exceed 10", using the value of the size of the Earth's orbit specified by various Copernicans.

| A | B | C | D | E | F | G |
|---|---|---|---|---|---|---|
| | | | \multicolumn{2}{c}{TRUE SIZE OF SIRIUS} | \multicolumn{2}{c}{TRUE SIZE OF ALCOR} |
| | | Required distances of stars from Earth (for parallax to be 10" or less), measured in Earth radii | Diameter (in Earth diameters) | Volume (in Earth volumes) | Diameter (in Earth diams) | Volume (in Earth volumes) |
| Copernicus | | 47,439,800 | 4170 | 71,667,713,000 | 1992 | 4,378,454,048 |
| Herigonius | | 49,502,400 | 4350 | 82,312,875,000 | 2068 | 8,844,058,432 |
| Galilaeus | | 49,832,416 | 4380 | 8,427,672,000 | 2092 | 9,155,562,688 |
| Bullialdus | | 60,227,920 | 5300 | 148,877,000,000 | 2530 | 15,941,277,000 |
| Lansbergius | | 61,616,122 | 5424 | 159,371,956,024 | 2588 | 17,333,761,472 |
| Keplerus | | 142,746,428 | 12550 | 1,976,656,375,000 | 6000 | 216,000,000,000 |
| Vendelinus | | 604,589,312 | 53200 | 15,056,882,800,000 | 25380 | 1,767,384,872,000 |
| \multicolumn{7}{c}{For the basis of these distances see Book 6, Chapter 7, Number 15} |

| \multicolumn{7}{l}{*With volumes calculated from Riccioli diameter values:*} |
|---|---|---|---|---|---|---|
| Copernicus | | 47,439,800 | 4170 | 72,511,713,000 | 1992 | 7,904,383,488 |
| Herigonius | | 49,502,400 | 4350 | 82,312,875,000 | 2068 | 8,844,058,432 |
| Galilaeus | | 49,832,416 | 4380 | 84,027,672,000 | 2092 | 9,155,562,688 |
| Bullialdus | | 60,227,920 | 5300 | 148,877,000,000 | 2530 | 16,194,277,000 |
| Lansbergius | | 61,616,122 | 5424 | 159,572,865,024 | 2588 | 17,333,761,472 |
| Keplerus | | 142,746,428 | 12550 | 1,976,656,375,000 | 6000 | 216,000,000,000 |
| Vendelinus | | 604,589,312 | 53200 | 150,568,768,000,000 | 25380 | 16,348,384,872,000 |

| \multicolumn{7}{l}{*With diameters and volumes calculated based on a maximum 10" parallax, on selected astronomers' values for radius of the Earth's orbit (in Earth radii; these values are listed by each name in Column B; values for Copernicus and Galileus are from Galileo and Drake [1967: 487, note p. 359]; for Keplerus and Vendelius are from Hooke [1705: 495]), and on the sizes of Sirius and Alcor:*} |
|---|---|---|---|---|---|---|
| Copernicus | 1179 | 48,637,241 | 2122 | 9,557,821,833 | 519 | 139,604,509 |
| Galilaeus | 1208 | 49,833,577 | 2174 | 10,280,596,599 | 532 | 150,161,581 |
| Keplerus | 3381 | 139,476,262 | 6086 | 225,399,541,149 | 1488 | 3,292,255,575 |
| Vendelinus | 14600 | 602,293,234 | 26280 | 18,149,977,152,000 | 6424 | 265,104,193,024 |

| With diameters and volumes calculated as above but in units of Earth radii: | | | | | | |
|---|---|---|---|---|---|---|
| | | | Diameter (in Earth radii) | | Diameter (in Earth radii) | |
| Copernicus | | 48,637,241 | 4244 | 76,462,574,664 | 1038 | 1,116,836,070 |
| Galilaeus | | 49,833,577 | 4349 | 82,244,772,790 | 1063 | 1,201,292,648 |
| Keplerus | | 139,476,262 | 12172 | 1,803,196,329,190 | 2975 | 26,338,044,600 |
| Vendelinus | | 602,293,234 | 52560 | 145,199,817,216,000 | 12848 | 2,120,833,544,192 |

| With diameters and volumes calculated as above but in units of Earth radii with factor of two error on Alcor: | | | | | | |
|---|---|---|---|---|---|---|
| | | | Diameter (in Earth radii) | | Diameter (in Earth radii) x 2 | |
| Copernicus | | 48,637,241 | 4244 | 76,462,574,664 | 2075 | 8,934,688,560 |
| Galilaeus | | 49,833,577 | 4349 | 82,244,772,790 | 2126 | 9,610,341,187 |
| Keplerus | | 139,476,262 | 12172 | 1,803,196,329,190 | 5951 | 210,704,356,798 |
| Vendelinus | | 602,293,234 | 52560 | 145,199,817,216,000 | 25696 | 16,966,668,353,536 |

Table 4: Riccioli's table (highlighted) from the *Almagestum Novum* of calculated physical diameters and volumes of stars, compared to the Earth, based on stars having an annual parallax of 10" or less, and on values for the radius of Earth's orbit provided by various astronomers. Calculations based on Riccioli's numbers are included for purpose of comparison, illustrating errors similar to those found in Table 2 and Table 3, as well as factor-of-ten errors in Galileus (column E) and Vendelius (columns E and G). These latter errors, especially Galileus E, make it clear that this part of the *Almagestum Novum* received very little proofreading.